\documentclass[12pt]{article}

\ifx\pdfoutput\undefined
\usepackage[dvips,bookmarks]{hyperref}
\else
\usepackage{hyperref}
\fi
\hypersetup{colorlinks=false,bookmarksopen,bookmarksnumbered,citecolor=blue,
   pdfstartview=FitH}

\usepackage{latexsym}

\usepackage{amssymb,amsfonts,amsmath}
\usepackage{graphicx} 
\usepackage{indentfirst}

 \usepackage{bbm}

\parskip=\medskipamount

\newcommand {\cN}{{\cal N}}

\newcommand {\cV}{{\cal V}}

\def\a{\alpha}

\def\b{\beta}

\def\d{\delta}

\def\g{\gamma}
\def\G{\Gamma}

\def\x{\xi}
\def\z{\zeta}

\def\J{\Psi}

\newcommand{\ad}{{\dot{\alpha}}}                           %new
\newcommand{\bd}{{\dot{\beta}}}                            %new
\newcommand{\pa}{\partial}                           %new
\newcommand{\hf}{\frac12}
\newcommand{\be}{\begin{equation}}
\newcommand{\ee}{\end{equation}}
\newcommand{\bea}{\begin{eqnarray}}
\newcommand{\eea}{\end{eqnarray}}
\newcommand{\non}{\nonumber}
\newcommand{\ba}{\begin{array}}
\newcommand{\ea}{\end{array}}

\newcommand{\bm}[1]{\mbox{\boldmath$#1$}}

\def\double #1{#1{\hbox{\kern-2pt $#1$}}}

\newcommand{\gd}{{\dot\g}}

\newcommand{\bsubeq}{\begin{subequations}}
\newcommand{\esubeq}{\end{subequations}}

\begin{document}

\begin{titlepage}

\begin{flushright}
March, 2011
\end{flushright}
\vspace{5mm}

\begin{center}
{\Large\bf  Comments on the gauge models for 
massless higher spin off-shell supermultiplets
}
\end{center}

\begin{center}

{\large 
Sergei M. 
Kuzenko\footnote{kuzenko@cyllene.uwa.edu.au}
}
\vspace{2mm}

\footnotesize{
{\it School of Physics M013, The University of Western Australia,\\
35 Stirling Highway, Crawley W.A. 6009, Australia}} 
\vspace{2mm}

\end{center}
\vspace{5mm}

\begin{abstract}
\baselineskip=14pt
In recent papers arXiv:1103.3564 and arXiv:1103.3565,
Gates and Koutrolikos announced the construction of new 
off-shell formulations
for massless higher spin  supermultiplets. Here we demonstrate that 
all of their models are obtained from (some of) those constructed in 1993 by 
Kuzenko,  Postnikov and Sibiryakov by applying special field redefinitions.
\end{abstract}

\vfill
\end{titlepage}

\newpage
\setcounter{page}{1}
\renewcommand{\thefootnote}{\arabic{footnote}}
\setcounter{footnote}{0}

\section{Introduction}
\setcounter{equation}{0}
In four space-time dimensions, the off-shell formulations
for massless higher spin  $\cN=1$ supermultiplets
were constructed in  \cite{KSP,KS}. For each superspin $s \geq 1$, 
half-integer \cite{KSP} and integer \cite{KS},
these publications provided two dually equivalent off-shell
realizations in ${\cal N }= 1$ Minkowski superspace.
At the component level, each of the
two superspin-$s$ actions \cite{KSP,KS} reduces, 
{\it upon} imposing a Wess-Zumino-type
gauge and eliminating the auxiliary fields,
to a sum of the spin-$s$ and  spin-$(s+1/2)$ actions \cite{Fronsdal1}.\footnote{The results obtained
in  \cite{KSP,KS} are reviewed in \cite{BK}.}
The higher spin superfield theories  of \cite{KSP,KS} were generalized to the case of ant-de Sitter supersymmetry 
in \cite{KS2}.

Recently, Gates and Koutrolikos \cite{GK1,GK2} have announced the construction of new 
off-shell formulations for massless higher spin  supermultiplets. 
In the present note  we demonstrate that  these models are, in fact, not new. 
They can be obtained from (some of) those constructed in \cite{KSP,KS}
by applying special field redefinitions. 

In the remainder of this section, we recall the structure of 
the constrained superfields used in \cite{KSP,KS}.
Then, in section 2 we consider the case of half-integer superspins.  Finally, 
section 3 is devoted to the case of integer-superspin models.

The off-shell 
formulations 
for massless higher spin 
massless supermultiplets developed in \cite{KSP,KS}
are realized in 4D $\cN=1$ 
Minkowski superspace.\footnote{Our superspace 
notation and conventions correspond to \cite {BK}, 
in particular the flat superspace
covariant derivatives are
$D_A = (\pa_a, D_\a,\bar D^\ad)$. 
Throughout
this paper we consider only Lorentz 
tensors  symmetric in their undotted indices and separately 
in their dotted ones.   
${}$For a tensor of type $(k,l)$   with $k$ undotted 
and $l$ dotted indices we use the shorthand notations
$ 
\Psi_{\a(k) \ad(l)}   \equiv \Psi_{\a_1 \ldots
\a_k\ad_1\ldots \ad_l} = \Psi_{(\a_1 \ldots \a_k)(\ad_1\ldots
\ad_l)}$.
Quite often
we assume that the upper or lower indices,
which are denoted by  one and the same letter,
should be symmetrized, for instance
$
\phi_{\a(k)} \psi_{\a(l)}
\equiv \phi_{(\a_1\ldots\a_k} \psi_{\a_{k+1}\ldots\a_{k+l})}$.
Given two tensors of the same type, their contraction is denoted by
$f \cdot g \equiv f^{\a(k) \ad(l)} \,g_{\a(k) \ad(l)}
= $    $f^{\a_1 \ldots \a_k\ad_1\ldots \ad_l}\, g_{\a_1 \ldots
\a_k\ad_1\ldots \ad_l}$.
} 
They involve the so-called transverse and longitudinal 
linear superfields, both as dynamical variables and gauge parameters. 
A complex tensor superfield $\G_{\a(k) \ad(l)}$ is said to be {\it transverse linear} if 
it obeys the constraint
\bea
{\bar D}^\bd \,\G_{\a(k) \bd \ad(l-1)} & = & 0 ~, \qquad l>0 ~.
\label{transverse}
\eea
A {\it longitudinal linear} superfield
$G_{\a(k) \ad(l)}$  is defined to satisfy the constraint
\be
{\bar D}_{ (\ad_1} \,G_{\a(k) \ad_2 \dots \ad_{l+1})}=0 ~.
\label{longit}
\ee
The above constraints imply that
$\G_{\a(k) \ad(l)}$ and $G_{\a(k) \ad(l)}$ 
are linear in the usual sense
\be
{\bar D}^2\,  \G_{\a(k) \ad(l)} = {\bar D}^2 \,G_{\a(k) \ad(l)} = 0 ~.
\label{5}
\ee
In the case $l=0$, the constraint  (\ref{transverse})
should be  replaced by ${\bar D}^2 \G_{\a(k) } =0$. The constraint 
(\ref{longit}) for $l=0 $ simply means that $G_{\a(k)} $ is chiral, 
${\bar D}_{ \bd} \,G_{\a(k) }=0$.
The constraints (\ref{transverse}) and 
(\ref{longit}) can be solved in terms of
unconstrained prepotentials
$ \Phi_{\a(k)\ad(l+1)} $ and $ \Psi_{\a(k)\ad(l-1)} $
as follows:
\begin{subequations}
\bea
 \G_{\a(k) \ad(l)}&=& \bar D^\bd 
{\bar \Phi}_{\a(k)\,\bd \ad_1 \cdots \ad_l } ~,
\label{tr-prep}
 \\
 G_{\a(k) \ad(l)} &=& {\bar D}_{( \ad_1 }
 \Psi_{ \a(k) \, \ad_2 \cdots \ad_{l}) } ~.
\label{7}
\eea
\end{subequations}
The prepotentials  are defined modulo gauge transformations of the form:
\begin{subequations}
\bea
\d {\bar \Phi}_{\a(k)\, \ad_{(l+1 )}} 
&=&  \bar D^\bd 
{\bar \x}_{\a(k)\, (\bd \ad_1 \cdots \ad_{l+1}) } ~,
\label{tr-prep-gauge}
\\
\d  \Psi_{ \a(k) \, \ad_{(l-1}) } &=&  {\bar D}_{( \ad_1 }
 \z_{ \a(k) \, \ad_2 \cdots \ad_{l-1}) } ~,
\label{lon-prep-gauge}
\eea
\end{subequations}
with the gauge parameters $ {\bar \x}_{\a(k)\,  \ad (l+2) } $
and $ \z_{ \a(k) \, \ad (l-2)}$ being unconstrained. 
In other words, the variations $\d {\bar \Phi}_{\a(k)\, \ad_{(l+1 )}} $ and 
$\d  \Psi_{ \a(k) \, \ad_{l-1}) } $ are transverse linear and longitudinal linear, respectively.

\section{Half-integer superspin}
\setcounter{equation}{0}

Two formulations for the massless multiplet of 
a half-integer superspin 
$s+1/2$ (with $s=1,2\ldots$) which were called 
in Ref. \cite{KSP} transverse and longitudinal, 
contain the following dynamical variables respectively:
\bea
\cV^\bot_{s+1/2}& = &\Big\{H_{\a(s)\ad(s)}~, ~
\G_{\a(s-1) \ad(s-1)}~,
~ \bar{\G}_{\a(s-1) \ad(s-1)} \Big\} ~,    \\
\label{10}
\cV^{\|}_{s+1/2} &=& 
\Big\{H_{\a(s)\ad(s)}~, ~
G_{\a(s-1) \ad(s-1)}~,
~ \bar{G}_{\a(s-1) \ad(s-1)} \Big\}
~.
\label{11}
\eea
Here $H_{\a(s) \ad (s)}$ is real, 
$\G_{\a (s-1) \ad (s-1)} $ transverse linear and 
$G_{\a (s-1) \ad (s-1)}$ 
longitudinal linear  superfields.  
The  case $s=1$  corresponds to 
linearized supergravity  
(see  \cite{BK} for  a review).  

The gauge transformations for the superfields 
$H_{\a(s) \ad (s)} $, $\G_{\a (s-1) \ad(s-1)}$
and $G_{\a (s-1) \ad (s-1)}$  
postulated in \cite{KSP}
are 
\bea
\d  H_{\a(s) \ad (s)}  &=& g_{\a(s) \ad (s)}  
+ {\bar g}_{\a(s) \ad (s)}  ~,  
\label{16}\\
\d  \G_{\a (s-1) \ad(s-1)}
&=& \hf \, \frac{s}{s+1} \, \bar D^\bd D^\b 
{\bar g}_{\b \a(s-1) \bd \ad(s-1)}~,
\label{17}\\
\d G_{\a (s-1) \ad (s-1)}
&=& \hf \, \frac s{s+1} \, 
D^\b \bar D^\bd 
g _{\b \a(s-1) \bd \ad(s-1)} + {\rm i} \,s\,
 \pa^{\b \bd } g _{\b \a(s-1) \bd \ad(s-1)} ~, 
\label{18}
\eea
with the gauge  parameter $g_{\a(s) \ad (s)}  $  being an arbitrary {\it longitudinal linear} 
superfield.
It can be seen that $\d G_{\a (s-1) \ad (s-1)}$ is longitudinal
linear.  
%Eq. (\ref{hi-t}) defines the action invariant 
%under the gauge transformations (\ref{16}) and
%(\ref{17}).  Similarly, eq. (\ref{hi-l}) defines  the action invariant 
%under the gauge transformations (\ref{16}) and (\ref{18}). 

In the transverse formulation, 
the action  invariant 
under the gauge transformations (\ref{16}) and
(\ref{17})  is 
\bea
S^{\bot}_{s+1/2}&=&
\Big( - \frac{1}{2}\Big)^s  \int {\rm d}^8z \,
\Big\{ \frac{1}{8} H^{ \a(s) \ad(s) }  D^\b {\bar D}^2 D_\b 
H_{\a(s) \ad(s) }  \non \\
&+& H^{ \a(s) \ad(s) }
\left( D_{\a_s}  {\bar D}_{\ad_s} \G_{\a(s-1) \ad(s-1) }
- {\bar D}_{\ad_s}  D_{\a_s} 
{\bar \G}_{\a (s-1) \ad (s-1) } \right) \non \\
&+&\Big( {\bar \G} \cdot \G
+ \frac{s+1} {s} \, \G \cdot \G ~+~ {\rm c.c.} \Big)
\Big\} ~.
\label{hi-t}
\eea
In the longitudinal formulation, the action  invariant 
under the gauge transformations (\ref{16}) and
(\ref{18}) 
is
\bea
S^{\|}_{s+1/2}&=&
\Big(-\hf \Big)^s  \int {\rm d}^8z \, \Big\{
\frac 18 
H^{\a (s) \ad (s) }   D^\b \bar D^2
 D_\b H_{\a (s) \ad (s) }  \non \\
&-& \frac{1}{8} \, \frac{s}{2s+1} \, \Big( \,
\big[ D_{\g}, \bar D_{\gd}\big] H^{\g \a (s-1)\gd  \ad (s-1)}
\, \Big)  \,
\big[ D^\b, \bar D^{\bd}\big]
H_{\b\a(s-1)\bd\ad(s-1)} \,  \non \\
&+& \frac{s}{2}\, \Big( \partial_{\gd} 
H^{\g \a (s-1) \gd \ad (s-1)}  \Big) \,
\partial^{\b\bd}
H_{\b\a(s-1)\bd\ad(s-1)} 
\non \\
&+& 2{\rm i} \, \frac{ s}{2s+1}  \,  \pa_{\g \gd } 
H^{\g \a (s-1) \gd \ad (s-1)}
\Big( G_{\a(s-1) \ad(s-1)} - \bar G_{\a(s-1) \ad(s-1)} \Big)  \non \\
&+& \frac{1}{2s+1} \Big( \bar G \cdot G - \frac{s+1}s G \cdot G
+ {\rm c.c.}\Big)\Big\}   ~.
\label{hi-l}
\eea
The models (\ref{hi-t}) and (\ref{hi-l}) are dually equivalent 
\cite{KSP}.

It was pointed out in \cite{KS2} that there is a natural freedom in the definition of 
$ \G_{\a (s-1) \ad(s-1)}$. Specifically, instead of working with $ \G_{\a (s-1) \ad(s-1)}$
one can introduce the following transverse linear superfield
\bea
{\bm  \G}_{\a (s-1) \ad(s-1)}:= \G_{\a (s-1) \ad(s-1)}
+c\, {\bar D}^\bd D^\b  H_{\b \a (s-1) \bd \ad(s-1)}~, 
\eea
with $c$ an arbitrary constant.
As follows from (\ref{16}) and
(\ref{17}), the gauge transformation law of ${\bm  \G}_{\a (s-1) \ad(s-1)}$ is 
\bea 
\d {\bm  \G}_{\a (s-1) \ad(s-1)}= 
\Big(c + \hf \, \frac{s}{s+1} \Big) \bar D^\bd D^\b 
{\bar g}_{\b \a(s-1) \bd \ad(s-1)} + c   \, \bar D^\bd D^\b 
{g}_{\b \a(s-1) \bd \ad(s-1)}~.
\label{2.9}
\eea
Clearly, the transverse theory (\ref{hi-t}) can be re-formulated in terms of 
${\bm  \G}_{\a (s-1) \ad(s-1)}$  and its conjugate \cite{KS2}. However, only in the case 
$c=0$, the action functional has the simplest form (\ref{hi-t}).

Now, choosing $c= - \hf \, \frac{s}{s+1}$  in (\ref{2.9}) gives
\bea 
\d {\bm  \G}_{\a (s-1) \ad(s-1)}= 
-\hf \, \frac{s}{s+1}    \, \bar D^\bd D^\b 
{g}_{\b \a(s-1) \bd \ad(s-1)}~.
\label{2.10}
\eea
This is exactly the novel transformation law introduced in \cite{GK1}.
More precisely, 
one has to fill in  a couple of technical details in order to see that the transformation 
(\ref{2.10}) indeed coincides with that advocated in \cite{GK1}, that is eq. (33) in \cite{GK1}. First, 
one has to express ${\bm \G}_{\a(s-1) \ad(s-1)}$ in terms of its prepotential,
${\bar {\bm \Phi}}_{\a(s-1)\, \ad (s) } $,  
in accordance with eq. (\ref{tr-prep}), which is defined modulo the {\it pre}-gauge 
transformations (\ref{tr-prep-gauge}). The point is that the model (32) introduced in \cite{GK1} 
is formulated in terms  of ${H}_{\a(s)\, \ad (s) } $, ${\bar {\bm \Phi}}_{\a(s-1)\, \ad (s) } $ and its conjugate 
${\bm \Phi}_{\a(s)\, \ad (s-1) } $.
Eq. (\ref{2.10}) leads  to the well-defined gauge transformation of ${\bar {\bm \Phi}}_{\a(s-1)\, \ad (s) } $. 
Secondly, in order to make a direct contact with \cite{GK1}, one should also represent the 
longitudinal linear parameter $g_{\a (s) \ad (s) }$  in eq. (\ref{2.10})
in the form  $g_{ \a (s) \ad (s)} = {\bar D}_{ ( \ad_1 } L_{ \a(s) \, \ad_2 \cdots \ad_s ) }$, 
for some unconstrained superfield $L_{\a (s) \ad (s-1) } $.
As a result, the complete gauge transformation of ${\bar {\bm \Phi}}_{\a(s-1)\, \ad (s) } $ is
\bea
\d {\bar {\bm \Phi}}_{\a(s-1)\, \ad (s) } = -\hf \, \frac{s}{s+1}    \,  D^\b 
{\bar D}_{(\ad_1} L_{\b \a(s-1) \ad_2 \cdots \ad_s )}
+ \bar D^\bd 
{\bar \x}_{\a(s)\, (\bd \ad_1 \cdots \ad_{s}) } ~,
\eea
which is the complex conjugate of the gauge transformation law  (33) in \cite{GK1}.

As a consequence of the above discussion, 
we conclude that the final gauge-invariant action given by  Gates and Koutrolikos, 
eq. (32) in \cite{GK1}, 
is obtained from (\ref{hi-t})
by applying the field redefinition expressing ${ \G}_{\a (s-1) \ad(s-1)}$ 
in terms of ${\bm  \G}_{\a (s-1) \ad(s-1)}$.

\section{Integer superspin}
\setcounter{equation}{0}

Two formulations of Ref. \cite{KS} 
for the massless multiplet of an integer superspin $s$ 
(with $s=1,2,\ldots$),  transverse and longitudinal,
contain the following dynamical variables  respectively:
\bea
\cV^\bot_s &=&
\Big\{H_{\a(s-1)\ad(s-1)}~, ~
\G_{\a(s) \ad(s)}~,
~ \bar{\G}_{\a(s) \ad(s)} \Big\} ~,    \\
\label{12}
\cV^\|_s &=&
\Big\{H_{\a(s-1)\ad(s-1)}~, ~
G_{\a(s) \ad(s)}~,
~ \bar{G}_{\a(s) \ad(s)} \Big\} ~.
\label{13}
\eea
Here $H_{\a(s-1) \ad (s-1)}$ is real, 
$\G_{\a (s) \ad (s)} $ transverse linear and 
$G_{\a (s) \ad (s)}$ 
longitudinal linear tensor superfields.
The case $s=1$
corresponds to the gravitino multiplet 
(see \cite{BK} for  a review).

The gauge transformations  
for the superfields $H_{\a (s-1) \ad (s-1)}$, 
$G_{\a (s) \ad(s)}$ and $\G_{\a(s) \ad (s)}$ 
postulated in \cite{KS} are
\bea
 \d   H_{\a (s-1) \ad (s-1)}&=& \g_{\a (s-1) \ad (s-1)} 
+ {\bar \g}_{\a (s-1) \ad (s-1)}~,
\label{21}\\
 \d  \G_{\a (s) \ad(s)}
 &=& \hf D_{(\a_s} {\bar D}_{(\ad_s} \,
 \g_{\a_1 \dots \a_{s-1} ) \ad_1 \dots \ad_ {s-1})}
-{\rm i}\, s \, \pa_{(\a_s (\ad_s } \,
\g_{\a_1 \dots \a_{s-1} ) \ad_1 \dots \ad_ {s-1})} ~, 
\label{22} \\
\d  G_{\a (s) \ad(s)} &=& 
\hf \bar D_{(\ad_s}  D_{( \a_s} \,
{\bar \g}_{ \a_1 \dots \a_{s-1} ) \ad_1 \dots \ad_ {s-1}) }~,
\label{23}
\eea
with the gauge parameter $\g_{\a(s-1) \ad(s-1)}$ being an arbitrary  {\it transverse linear} superfield.
It can be seen that $\d \G_{\a (s) \ad (s)}$ is transverse
linear.

In the transverse formulation, the action invariant  under the gauge transformations (\ref{21}) and (\ref{22})
is as follows:
\bea
S^{\bot}_s&=&- \Big(-\hf\Big)^s  \int {\rm d}^8z\, \Big\{-
\frac{1}{8} H^{\a(s-1)\ad(s-1)}  D^\b \bar D^2
 D_\b H_{\a(s-1)\ad(s-1)}  \non \\
&+& \frac{1}{8} \, \frac{s^2}{(s+1)(2s+1)} \,
 \Big(  \big[ D^{\a_s}, \bar D^{\ad_s}\big] 
H^{\a (s-1) \ad(s-1)} \Big) \, 
\big[ D_{(\a_s}, \bar D_{(\ad_s}\big] 
H_{\a_1 \dots \a_{s-1}) \ad_1 \dots \a_{s-1})} \,  \non \\
&+& \hf \, \frac{s^2}{s+1} \,
 \Big( \pa^{\a_s \ad_s}  H^{\a(s-1) \ad (s-1)} \Big)\,
 \pa_{(\a_s (\ad_s}  H_{\a_1 \dots \a_{s-1}) \, 
\ad_1 \dots \ad_{s-1})}   \non \\
&+& 2{\rm i}\, \frac{s}{2s+1} \, H^{\a(s-1) \ad (s-1) }
\pa^{\a_s \ad_s} 
\Big( \G_{ \a(s) \ad(s) } - \bar\G_{\a(s) \ad(s)} \Big) \non \\
&+& \frac{1}{2s+1} \, \Big( \bar\G \cdot \G
-  \frac{s+1}{s} \, \G \cdot \G + {\rm c.c.} \Big) \Big\}~.
\label{i-t}
\eea
In the longitudinal formulation, 
the action invariant  under the gauge transformations (\ref{21}) and (\ref{23})  is
\bea
S^{\|}_s &=&
\Big(-\hf  \Big)^s \int {\rm d}^8z \, \Big\{
\frac18 H^{\a(s-1) \ad(s-1)}  
D^\b \bar D^2 D_\b H_{\a(s-1) \ad(s-1) } \non \\
&+& \frac{s}{s+1}  H^{\a(s-1) \ad(s-1)} 
\left( D^\b  \bar D^\bd G_{\b \a (s-1) \bd \ad (s-1)} -
\bar D^{\bd} D^{\b} 
\bar G_{\b\a(s-1) \bd \ad (s-1)} \right) \non \\
&+&\Big( \bar G \cdot G
+  \frac{s}{s+1}\, G \cdot G
+ {\rm c.c.}\Big)\Big\} ~.
\label{i-l}
\eea
The models (\ref{i-l}) and (\ref{i-t}) are dually equivalent 
\cite{KS}.

It was pointed out in \cite{KS2} that there is a natural freedom in the definition of 
$ G_{\a (s) \ad(s)}$. Specifically, instead of working with $ G_{\a (s) \ad(s)}$
one can introduce the following longitudinal  linear superfield
\bea
{\bm  G}_{  \a (s) \ad(s)}:= G_{\a (s) \ad(s)}
+c\, {\bar D}_{( \ad_s} D_{(\a_s}  H_{ \a_1 \dots \a_{s-1} ) \ad_1 \dots \ad_ {s-1}) }~, 
\eea
with $c$ an arbitrary constant.
In accordance with  (\ref{21}) and (\ref{23}), the gauge transformation law of 
${\bm  G}_{  \a (s) \ad(s)}$ is 
\bea
\d {\bm  G}_{  \a (s) \ad(s)} = 
c  \bar D_{(\ad_s}  D_{( \a_s} \,
{ \g}_{ \a_1 \dots \a_{s-1} ) \ad_1 \dots \ad_ {s-1}) }
+(c + \hf ) \bar D_{(\ad_s}  D_{( \a_s} \,
{\bar \g}_{ \a_1 \dots \a_{s-1} ) \ad_1 \dots \ad_ {s-1}) }~.
\label{3.9}
\eea
Clearly, the longitudinal theory (\ref{i-l}) can be re-formulated in terms of 
${\bm  G}_{\a (s) \ad(s)}$  and its conjugate \cite{KS2}. However, only in the case 
$c=0$, the action functional has the simplest form (\ref{i-l}).

Now, choosing $c=-1/2$ in (\ref{3.9}) gives
\bea
\d {\bm  G}_{  \a (s) \ad(s)} = 
-\hf   \bar D_{(\ad_s}  D_{( \a_s} \,
{ \g}_{ \a_1 \dots \a_{s-1} ) \ad_1 \dots \ad_ {s-1}) }~.
\label{3.10}
\eea
This is exactly the novel transformation law introduced in \cite{GK2}. 
More precisely, 
one has to fill in  several technical details in order to see that the transformation 
(\ref{3.10}) indeed coincides with that advocated in \cite{GK2}, 
that is eq. (27) in \cite{GK2}. First, 
one has to express ${\bm G}_{\a(s) \ad(s)}$ in terms of its prepotential,
${\bm \J}_{\a(s)\, \ad (s-1) } $,  
in accordance with eq. (\ref{7}), which is defined modulo the {\it pre}-gauge 
transformations (\ref{lon-prep-gauge}).
The point is that the model (37) introduced in \cite{GK2} 
is formulated in terms  of ${H}_{\a(s-1)\, \ad (s-1) } $, 
${\bm \J}_{\a(s)\, \ad (s-1) } $ and its conjugate 
${\bar {\bm \J}}_{\a(s-1)\, \ad (s) } $.
Eq. (\ref{3.10}) leads  to the well-defined gauge transformation of ${ {\bm \J}}_{\a(s)\, \ad (s-1) } $. 
 Secondly, 
in order to make a direct contact with \cite{GK1}, one should also represent the 
transverse  linear parameter $\g_{\a (s-1) \ad (s-1) }$  in eq. (\ref{3.10})
in the form  $\g_{ \a (s-1) \ad (s-1)} = {\bar D}^{ \bd } {\bar L}_{ \a(s-1) \, (\bd \ad_1 \cdots \ad_{s-1} ) }$, 
for some unconstrained superfield ${\bar L}_{\a (s-1) \ad (s) } $.
As a result, the complete gauge transformation of ${\bm \J}_{\a(s)\, \ad (s-1) } $ is
\bea
\d {\bm \J}_{\a(s)\, \ad (s-1) } = -\hf D_{( \a_1} \,
{\bar D}^{ \bd } {\bar L}_{ \a_2 \cdots \a_{s} ) \, (\b \ad_1 \cdots \ad_{s-1} ) }
+  {\bar D}_{( \ad_1 }
 \z_{ \a(s) \, \ad_2 \cdots \ad_{s-1}) } ~,
\eea
which is the gauge transformation law  (27) in \cite{GK2}. 

As a consequence of  the above discussion, 
we conclude that the final gauge-invariant action given by  Gates and Koutrolikos,
eq. (37) in \cite{GK2}, 
is obtained from our action  (\ref{i-l})
by applying the field redefinition expressing ${ G}_{\a (s) \ad(s)}$ 
in terms of ${\bm  G}_{\a (s) \ad(s)}$.

It should be mentioned that Refs. \cite{GK1,GK2} presented interesting reformulations 
of the models studied, which involve an auxiliary unconstrained real superfield $B_{\a(s-1) \ad (s-1) }$. 
These reformulations may be useful and deserve further studies. 
\\

\noindent
{\bf Acknowledgements:} The author acknowledges email correspondence with Jim Gates.
This work  is supported in part by the Australian Research Council.

\footnotesize{

}

\end{document}